\documentclass[conference]{IEEEtran}
\IEEEoverridecommandlockouts
\usepackage{cite}
\usepackage{amsmath,amssymb,amsfonts}
\usepackage{algorithmic}
\usepackage{graphicx}
\usepackage{textcomp}
\usepackage{xcolor}
\usepackage{makecell}
\def\BibTeX{{\rm B\kern-.05em{\sc i\kern-.025em b}\kern-.08em
    T\kern-.1667em\lower.7ex\hbox{E}\kern-.125emX}}
    
\usepackage{enumerate}
\usepackage{tikz}
\usetikzlibrary{arrows,shapes.gates.logic.US,shapes.gates.logic.IEC,calc}
\usepackage{aurical}
\usepackage[T1]{fontenc}
\usepackage{xcolor}

\usepackage{amsmath}
\usepackage{subcaption}
\usepackage{graphicx}
\usepackage{color,soul}
 \usepackage{url}
\usepackage[normalem]{ulem}
\begin{document}

\title{Dynamic CU-DU Selection for Resource Allocation in O-RAN  Using Actor-Critic Learning}

\author{\IEEEauthorblockN{ Shahram Mollahasani$\dag$, Melike Erol-Kantarci$\dag$, Rodney Wilson$\ddag$}
\IEEEauthorblockA{\textit{$\dag$School of Electrical Engineering and Computer Science, University of Ottawa}\\
\textit{$\ddag$Ciena Corporation} \\
Emails: smollah2@uottawa.ca, melike.erolkantarci@uottawa.ca, rwilson@ciena.com}



}
\maketitle
\begin{abstract}
Recently, there has been tremendous efforts by network operators and equipment vendors to adopt intelligence and openness in the next generation radio access network (RAN). The goal is to reach a RAN that can self-optimize in a highly complex setting with multiple platforms, technologies and vendors in a converged compute and connect architecture. 
In this paper, we propose two nested actor-critic learning based techniques to optimize the placement of resource allocation function, and as well, the decisions for resource allocation. By this, we investigate the impact of observability on the performance of the reinforcement learning based resource allocation. We show that when a network function (NF) is dynamically relocated based on service requirements, using reinforcement learning techniques, latency and throughput gains are obtained.
\end{abstract}


\section{Introduction}
The fifth-generation (5G) mobile network has been designed to meet various requirements that can be mapped into three classes;  ultra-reliable and low-latency communications (URLLC), enhanced mobile broadband (eMBB), and massive machine-type communications(mMTC). However, the current cellular architecture suffers the lack of versatility and intelligence to satisfy these requirements efficiently \cite{z1,z2}. Therefore, an architectural reformation is required to be able to handle agility and heterogeneous technologies towards the evolution of the next generation of mobile networks (beyond 5G and 6G). One of the possible candidates that can enable the expected transformation is applying openness in the radio access network (RAN) along with intelligence, virtualization, and disaggregation. 
Open radio access network (O-RAN) alliance was formed in 2018, following a merger between the cloud RAN (C-RAN) alliance and xRAN forum to drive higher levels of openness. The primary purpose of O-RAN is to improve RAN utilization through the virtualization of network equipment and employing open interfaces to allow multiple vendors to operate and enhance RAN's intelligence. 

In O-RAN, the protocol stack can be divided into a centralized unit (O-CU) and a distributed unit (O-DU). For the sake of brevity, we will use CU and DU in the rest of the paper. The DU is located closer to the radio units (O-RU). The CU can be located where it can access higher processing power, such as data centers. The exact split of network functions between CU and DU is known as the functional split. By taking advantage of network function virtualization (NFV) techniques and running them at CUs, one can transfer part of network functions into data centers instead of running them at a base station (BS) \cite{z16}, which is a significant enabler for improving the NFVs' performance. The logical distribution of network functions (NFs) can be employed for different goals since they have various necessities and can be beneficial for the efficient management of network resources. This paper examines how by employing a reinforcement learning (RL) model, we can exploit the O-RAN architecture to improve an RL-based resource block allocation model's performance in a dynamic network.

In this paper, we aim to optimize the performance of an RL-based resource allocation model by selecting the best observation space using actor-critic learning. The observation of available resources has different availability at DU and CU. In addition, observability comes at a cost in terms of bandwidth requirement between CU-DU. We consider the packets' delay budget and traffic class priority of UEs and optimize the placement of the NF, i.e., decide to relocate it either at the DU or the CU dynamically. We employ various traffic types with different quality of service (QoS) requirements and show that by dynamically relocating an NF at different layers of O-RAN (in this case is RL-based resource block allocation model), the NF's performance (mean delay, packet delivery ratio, throughput) can be enhanced noticeably.

The main contributions of this work are as follows:
\begin{itemize}
\item We implement two RL models, one for allocating resource blocks to UEs by considering traffic types, delay budget, and their priorities, and the other one to taking advantage of the O-RAN architecture by distributing the processing load at different layers by considering available processing power and end-to-end delay in the network.
\item We improved the NF's performance by dynamically selecting between CU and DU to relocate the NF.
\end{itemize}

The rest of this paper is organized as follows: In section II, recent works are discussed. The system model is illustrated in section III. In section IV, the proposed RL-based model is comprehensively explained. In section V, the proposed model is examined, and it is compared with baseline models, and in section VI, we conclude the paper.

\section{related work}
One of the challenging issues which currently cellular operators are facing is achieving ubiquitous connectivity among network devices with heterogeneous QoS requirements \cite{c7}. In 5G, this problem becomes more severe since the network architecture is more complex and QoS requirements are more strict \cite{c8}. In addition, virtualization network functions need to dynamically share connect and compute resources to reach and maintain the required QoS for all UEs. To address the resource allocation problem many QoS-aware schedulers have been presented in the literature. In \cite{b1}, authors proposed a scheduler, which can guarantee the QoS requirements by various traffic types through encapsulating different traffic features for downlink scheduling. 
In \cite{b2}, a frame-level scheduler (FLS) is proposed, which prioritized real-time traffics with respect to the elastic ones such as file transfer or HTTP. 

RL-based techniques are also employed for optimizing various network functions in wireless networks \cite{z21}. In \cite{e3}, an RL-based resource block allocation model is proposed, which is applied in a reliable vehicle-to-vehicle (V2V) network. The presented RL agent learn and optimize the scheduler for allocating resource block to vehicles by frequently interacting with the environment. In \cite{e5}, an RL-based deadline-driven scheduler for transferring data is introduced. In this work, a central controller is considered, where requests will be scheduled based on their pacing rate. In \cite{e1}, authors propose a traffic predictor model that is trained to provide a reliable end-to-end RB allocation by considering dataset-dependent generalized service level agreement (SLA) constraints.

In \cite{z18}, RL is employed to optimize energy consumption and reduce the cost of a mobile network operator by dynamically applying function split in disaggregated RANs. Additionally, In \cite{z19}, tackle the user-cell association and beam allocation problem by introducing a transfer learning-based model tuned for 5G-NR mmWave networks. In this paper, we improve our previous work, an RL-based resource block allocation that allocates resource blocks to UEs by considering their channel condition, the priority of assigned packets, and the delay budget of UEs' traffic \cite{z20}. In \cite{z20}, only the scheduling problem is addressed. In this paper, we focus on joint CU-DU selection in O-RAN and resource allocation. 

Different than previous works, in this paper, we propose an actor-critic learning model that aims to optimize a network function's performance by selecting the best observation space. In the disaggregated RAN, the observation space can vary depending on the placement of the NF at the DU or CU. To the best of our knowledge, this problem has not been addressed in the literature before.

\begin{figure*}[]
\centering
\includegraphics [width=.75\linewidth] {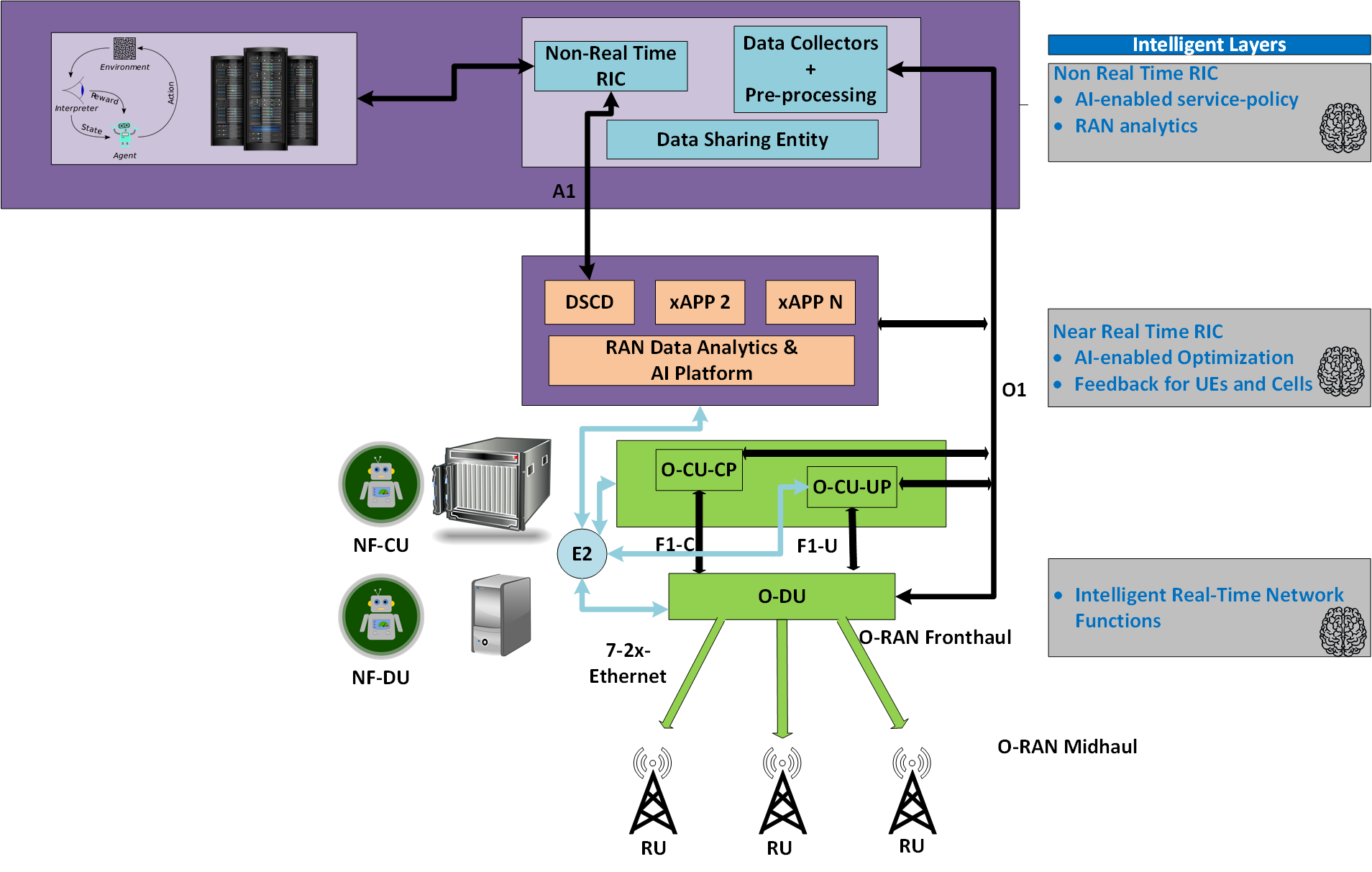} 
\caption{An overview of a dynamic CU-DU selection for resource allocation in O-RAN.}
\label{rl-network}
\end{figure*}

\section{System Model}
In this paper, we consider two nested RL agents which are tailored for the O-RAN architecture. The first one is an RL-based resource block allocation scheduler. The second one dynamically relocates the scheduler, with respect to traffic types and its delay budgets, at different layers of O-RAN. A system overview is illustrated in Fig. \ref{rl-network}. Accordingly, we assume there are \textit{N} UEs, \textit{M} BSs, and in order to reduce the observation space, the resource block group (RBG) is assigned as a unit for allocating RBs in the network. Additionally, we assume that each radio unit (RU) is connected to a distributed unit (DU), which is equipped with an edge server, while the central unit (CU) is located at a data center, far from BSs, with a higher processing power that can observe multiple DUs. Therefore, when an agent is located at the CU, it can access RUs’ resource block maps, avoid inter-cell interference, and apply more coordinated actions. The proposed actor-critic model aims to address the tradeoff of running the RL-based RB scheduler at DUs or at the CU by improving the reliability and satisfying the required delay budget for each traffic type.

\section{Dynamic CU-DU Selection Technique (DSCD)}
    In this model, the spatial pattern of traffic types and their requirements (delay budget, bit error rate, etc.) are learned by an actor-critic model. Additionally, to dynamically relocate the resource allocation function and improve the model's performance by providing higher observability, we employ a second RL-based algorithm that can distribute the loads based on the type of traffic and its priority at different layers of O-RAN. 

Although expanding the observation level can improve the model's accuracy, the complexity of the model will also increase, which can negatively affect the processing time. Moreover, the amount of time required by each agent to apply their actions (resource block allocation) can be varied with respect to the frequency, type, and delay budget of packets. Additionally, the availability of processing power at each layer of O-RAN architecture will be different. For instance, due to its limited processing power, DU is suitable for processing low complexity network functions. Meanwhile, since CUs are mainly deployed in data centers, where higher processing power is available, they are suitable for more complex algorithms that have larger state space. On the other hand, DUs are usually implemented near UEs; therefore, they can be a better option than a CU for relocating network functions serving traffic with tight delay requirements. Therefore, we need an intelligent algorithm that can perform dynamic relocating agents at either CU or DU to execute network functions by considering the availability of processing power, the complexity of the model, and its priority for applying actions with respect to the frequency type and delay budget of packets.

We employ an actor-critic learning model since it hierarchically provides control. In actor-critic learning techniques, value function and policy model are the main parts of the policy gradient. To decrease the gradient variables, the corresponding value function should be learned in order to assist and update the system policy. 
In the proposed model, based on the priority of traffic and its required delay budget, the corresponding location for the RL-scheduler agent is decided. Additionally, another RL-based scheduler allocates RBs to UEs by considering the channel quality indicator (CQI), the assigned traffic budget to each traffic, and the type of packets identified by the QoS Class Identifier (QCI). Suppose the scheduler agent can be processed at CU. In that case, it can access a wider range of observations from multiple DUs, and higher processing power is available for the agents. However, since some packets such as URLLCs are delay-sensitive, they need to be scheduled quickly, and transferring them to an AI agent located at CU will increase the processing time, which can amplify the packet drop rate. Therefore, in the proposed model, our aim is to assign a proper location to evaluate a network function by considering traffic type and their required delay budget. To do so, we employed a neural network (NN) that consists of three layers. During the learning procedure, neurons' weights are updated every transmission time interval (TTI) through the actor and the critic interactions. Similarly, the NN is tuned by the updated weights' value in the exploitation state, and it works as a non-linear function. RL agents try to increase the received reward at each state \textit{t}, which can be achieved by action-value $Q(s,a)$ and state-value $V(s)$ functions. The action-value function can be used for estimating the outcome of applying action \textit{a} in state \textit{s}. In contrast, the state-value function can be employed for estimating the average output of states. In this work, we use the Advantage Actor-Critic (A2C) model to reduce the number of parameters and simplify the learning stage. In the A2C model, instead of obtaining state-value and action-value functions, we just need to estimate $V(s)$. Additionally, in A2C, the advantage is a metric to examine the performed action with respect to the expected value $V(s)$ and it can be presented as $A(s_t,a_t)=Q(s_t,a_t)-V(s_t)$. Moreover, A2C, unlike asynchronous actor-critic (A3C), is a synchronous model with better consistency and can be a suitable candidate for disaggregated deployments \cite{c10}. The proposed model contains two neural networks (one for the actor and another for the critic):
\begin{itemize}
\item One of the NNs is utilized in the critic to estimate the value function, which can be used to criticize and enhance the actors' actions.
\item The other NN is used in agents, where the actions (choosing a proper location for the network function during each state) are applied in each time interval. 
\end{itemize}
In the following, we illustrated the structure of the proposed actor-critic model and its components in detail.
\begin{itemize}
\item Actor: The actor is responsible for exploring the corresponding policy $\pi_{\theta}$, which $\theta$ is the policy parameter, with respect to its observation ($O$) in order to apply a proper action ($A$).
\begin{equation}
\pi_{\theta }(O)=O\rightarrow A
\end{equation}
Consequently, the agent's action can be shown as:
\begin{equation}
a=\pi_{\theta }(O),
\end{equation}
here, $a\in A$. 
Actions in DSCD are defined as the location of executing the network function (CU or DU). Since actions are discrete, the softmax function is chosen as the last layer of the actor's NN to assign proper values for available actions in each state. The action's value shows the probability of obtaining a high reward by choosing that action, and the summation of the action's value is equivalent to 1.    

\item Critic: The critic is employed for estimating the value function $V(O)$. In each stat, actors are applying their actions, and then the critic obtain the temporal difference (TD) based on the next state ($O_{t+1}$) and the assigned reward value ($R_t$) as follows:
\begin{equation}
\delta _t= R_t+\gamma V (O_{t+1})- (O_{t}).
\end{equation} 
Here, TD error at time interval \textit{t} is presented as $\delta _t$, discount factor is $\gamma$. Additionally, to update the critic, the least-squares temporal difference (LSTD) should be minimized:
\begin{equation}
V^*=arg \ \underset{V}{min}(\delta _t)^2,
\end{equation}
here, $V^*$ is presented as the optimal value function. 
By using policy gradient, actors can be updated, and it can be achieved as follows:
\begin{equation}
\bigtriangledown_{\theta } J(\theta)=E_{\pi_{\theta }}[\bigtriangledown_{\theta }log {\pi_{\theta }}(O,a)\delta _t],
\end{equation}
here,  $\bigtriangledown_{\theta } J(\theta)$ and $\pi_{\theta }(O,a)$ are the gradient of the cost function and  the action's value under the current policy, respectively. Then at the actor, we can obtain the parameters' weight difference as follows:
\begin{equation}
 \Delta_{\theta_t}=\alpha \bigtriangledown_{\theta_t }log \pi_{\theta_t}(O_t,a_t)\delta _t,
\end{equation}
here, ($\bigtriangledown_{\theta_t }$) is the estimated gradient per time step, which will be used for updating the parameters. Moreover, $\alpha$ is the learning rate and it is between 0 and 1. Finally, the actor's network can be updated through policy gradient as follows: 
\begin{equation}
\theta_{t+1}=\theta_t+\alpha \bigtriangledown_{\theta_t }log \pi_{\theta_t}(O_t,a_t)\delta _t.
\end{equation}
\end{itemize}

Furthermore, agents  take actions based on the received reward in each state. In the RL-based scheduler the reward function is defined as follows \cite{z20}:

\begin{align}
R= R_1 + R_2 + R_3,\\
R_1= max\left ( sgn\left ( cqi_k-\frac{\sum_{j=0}^{K}cqi_j}{K} \right ),0 \right ) \label{r1}\\
R_2= \left\{\begin{matrix}
1 & Packet_{URLLC}\\ 
0 & Otherwise
\end{matrix}\right.\label{r2}\\
R_3=sinc(\pi \left \lfloor \frac{Packet_{delay}}{Packet_{budget}} \right \rfloor) \label{r3}
\end{align}
Here, The feedback of $UE_k$ is presented as $cqi_k$, $Packet_{URLLC}$ is for URLLC packets, while $ Packet_{delay}$ and $Packet_{budget}$ are HoL delay of packets and the delay budget assigned to each traffic types, respectively.

We define the reward for DSCD as:

\begin{align}
R=\tau (U \times D) +\lambda R_3,\\
R_3=sinc(\pi \left \lfloor \frac{Packet_{delay}}{Packet_{budget}} \right \rfloor) \label{r5}
\end{align}

Here, \textit{U} and \textit{D} are binary values and are equal to 1 when the traffic type is URLLC and NF is executed at DU, respectively. Based on the type of packets, which can be identified through the QoS Class Identifier (QCI) value, if the packet is a URLLC packet ($Packet_{URLLC}$) and it is processed at the DU, the agent will receive an extra reward. Additionally, we need to make sure the delay budge ($Packet_{budget}$) of other packets can be met. Therefore, if the packet's delay ($ Packet_{delay}$), which can be considered as the head of the line (HoL) delay (the time that packets remain in RLC buffer), is less than the predefined packets delay budget, agents will receive another extra reward. $\tau$ and $\lambda$ are scalar weights to maintain the URLLC load at DU and satisfy the delay budget for all traffics in each state. 

\section{Performance evaluation}
We implemented the proposed model using ns3-gym \cite{c11}. 
The NN of the proposed model is implemented in Pytorch. In simulations, we considered 4 BSs and 40 UEs, which are distributed randomly. The density of URLLC UEs varies between 10\% to 30\%. In this work, numerology zero with 12 subcarriers, which has 15 KHz subcarrier spacing and 14 symbols per subframe, is employed. Additionally, scheduling is done every 1 TTI. The presented results are an average of 10 runs with 5000 iterations.

The performance of the proposed model is examined in two scenarios. In the first case, UEs are fixed, and we have two traffic types, including live video stream and Augmented reality (AR). In the second case, we have mobility among UEs and vehicle-to-everything (V2X) traffic (defined with respect to 5GAA standards \cite{c14}).

\begin{figure*}[h]
       \begin{subfigure}[b]{0.23\textwidth}
         \centering
         \includegraphics[width=\textwidth]{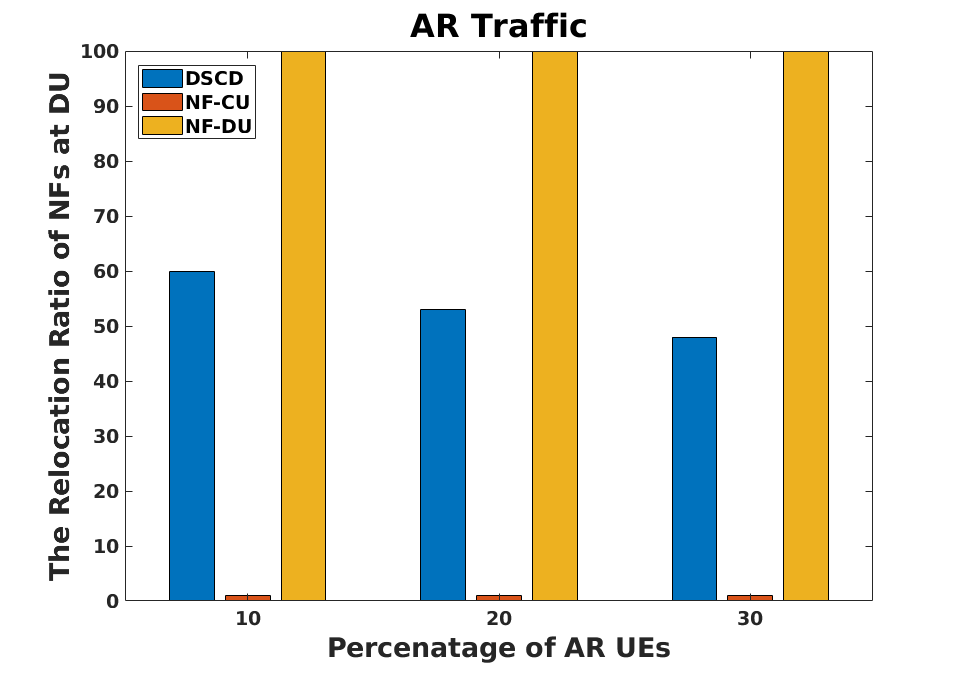}
         \caption{AR traffic's ratio at DU}
         \label{ratio_ur_du}
     \end{subfigure}
     \hfill
       \begin{subfigure}[b]{0.23\textwidth}
         \centering
         \includegraphics[width=\textwidth]{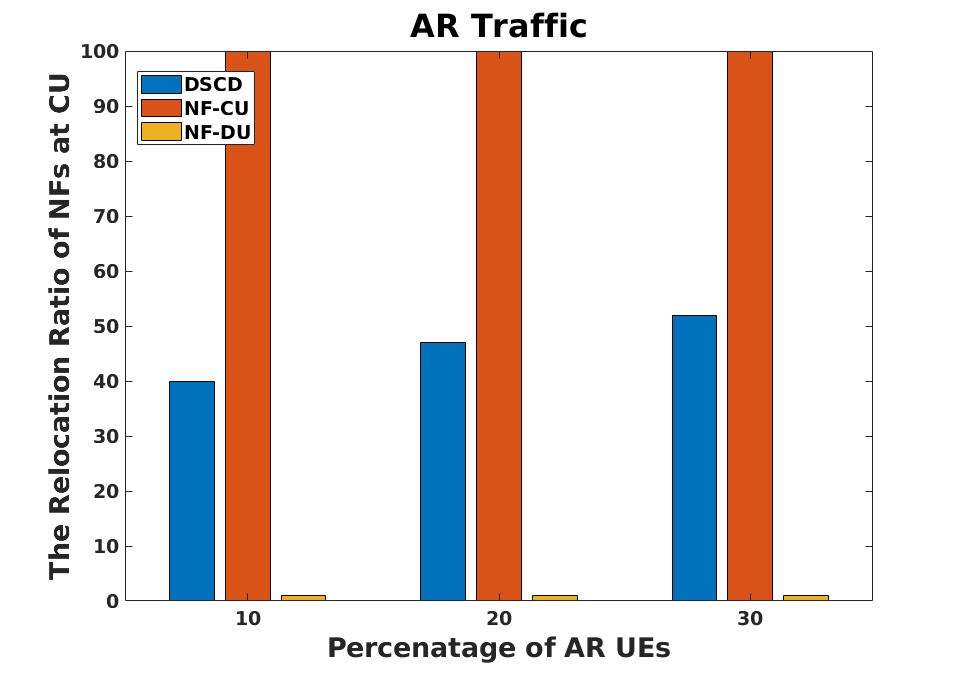}
         \caption{AR traffic's ratio at CU}
         \label{ratio_ur_cu}
     \end{subfigure}
     \hfill
       \begin{subfigure}[b]{0.23\textwidth}
         \centering
         \includegraphics[width=\textwidth]{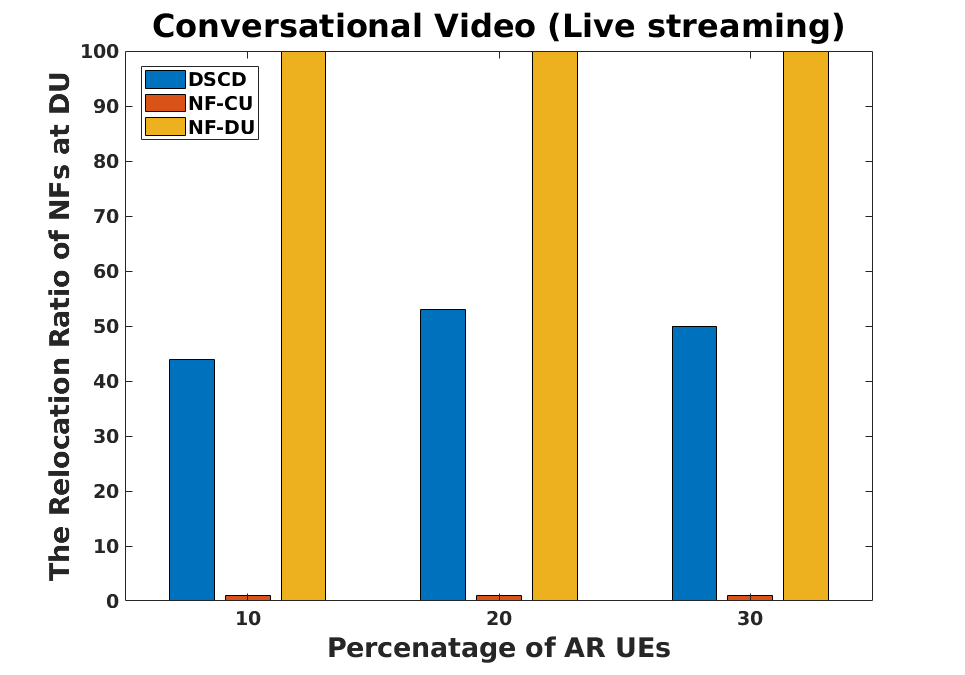}
         \caption{Video traffic's ratio at DU}
         \label{ratio_vd_du}
     \end{subfigure}
     \hfill
       \begin{subfigure}[b]{0.23\textwidth}
         \centering
         \includegraphics[width=\textwidth]{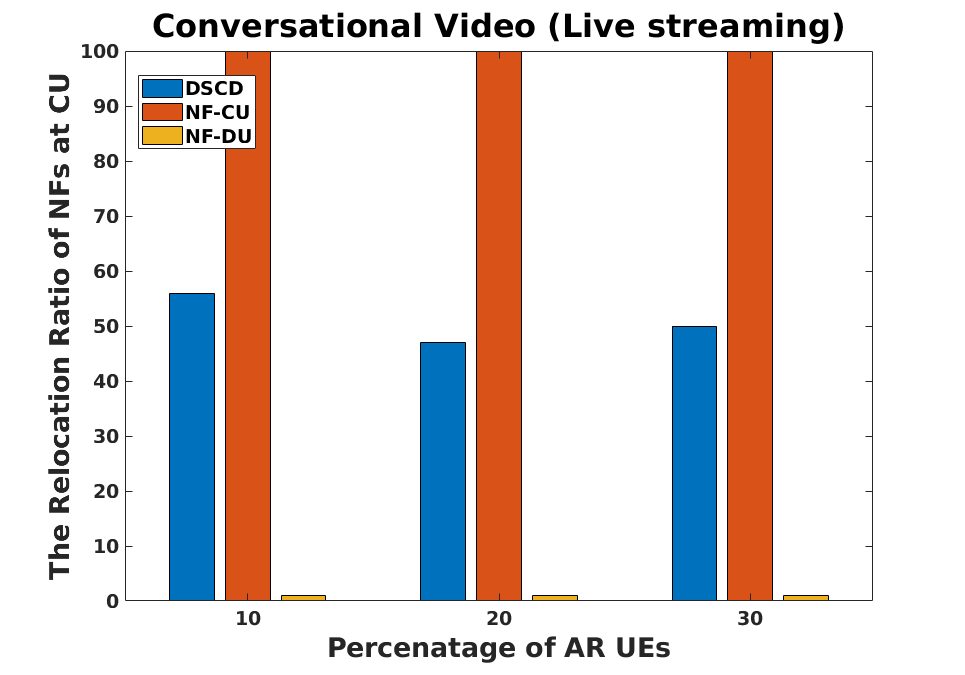}
         \caption{Video traffic's ratio at CU}
         \label{ratio_vd_cu}
     \end{subfigure}
     \hfill     
  \caption{The relocation ratio of the NF at DU and CU in the fixed scenario.}
  \label{loadfix}
\end{figure*}


Note that we assume that packets will be dropped if they stay in the RLC queue more than their delay budget, and they will be considered satisfied if the scheduler can assign their required RBs within their delay budget. In Table. I, the QoS metrics of traffic types, are presented.

\begin{table}[h]
\centering
\caption{Traffic properties \cite{b11}.}
\begin{tabular}{|l|l|l|l|l|}
\hline
QCI & \multicolumn{1}{c|}{\begin{tabular}[c]{@{}c@{}}Resource\\ Type\end{tabular}} & Priority & \begin{tabular}[c]{@{}l@{}}Packet\\ Delay\\ Budget\end{tabular} & \begin{tabular}[c]{@{}l@{}}Service \\  Example\end{tabular} \\ \hline
2   & GBR                                                                          & 40        & 150 ms                                                          & Live stream video                                                         \\ \hline
80   & Non-GBR                                                                      & 68        & 10 ms                                                          & AR                                                          \\ \hline
75  & GBR                                                                          & 25      & 20 ms                                                           & V2X                                                           \\ \hline
\end{tabular}
\label{packets}
\end{table}

In order to examine the performance of the proposed model, we consider two benchmark schemes, namely NF-DU in which DU always handles the NF, While in the other one, the NF will always be handled by the CU (named by NF-CU), which are developed based on a channel, delay and priority
aware actor-critic learning-based scheduler (CDPA-A2C) \cite{z20}.  In Table II. the simulation parameter and the configuration of NNs for the actor and critic learning are shown in detail. 
\begin{table}[h]
\centering
\caption{Simulation parameters.}
\begin{tabular}{|c|c|}
\hline
Parameters                                                                       & Value                           \\ \hline
Number of neurons                                                                & 900 x 100 (Actor + Critic) \\ \hline
Scheduler algorithm                                                              & CDPA-A2C                       \\ \hline
Number of BSs                                                                    & 4                               \\ \hline
Number of UEs                                                                    & 40                          \\ \hline
\begin{tabular}[c]{@{}c@{}}Maximum Traffic load per UE\\ (Downlink)\end{tabular} & 256 kbps                        \\ \hline
Traffic types                                                                    &  Video, VR, V2X          \\ \hline
Traffic stream per TTI                                                                    & 50          \\ \hline

DSCD reward's weights                                                                  & \makecell{ $\tau=0.5$ and $\lambda=0.5$ }  \\ \hline
Discount factor                                                                  & 0.9                             \\ \hline
Actor learning-rate                                                              & 0.01                            \\ \hline
Critic learning-rate                                                             & 0.05                            \\ \hline
\end{tabular}
\end{table}

\begin{figure}[h]
  \begin{subfigure}{\linewidth}
  \includegraphics[width=.5\linewidth]{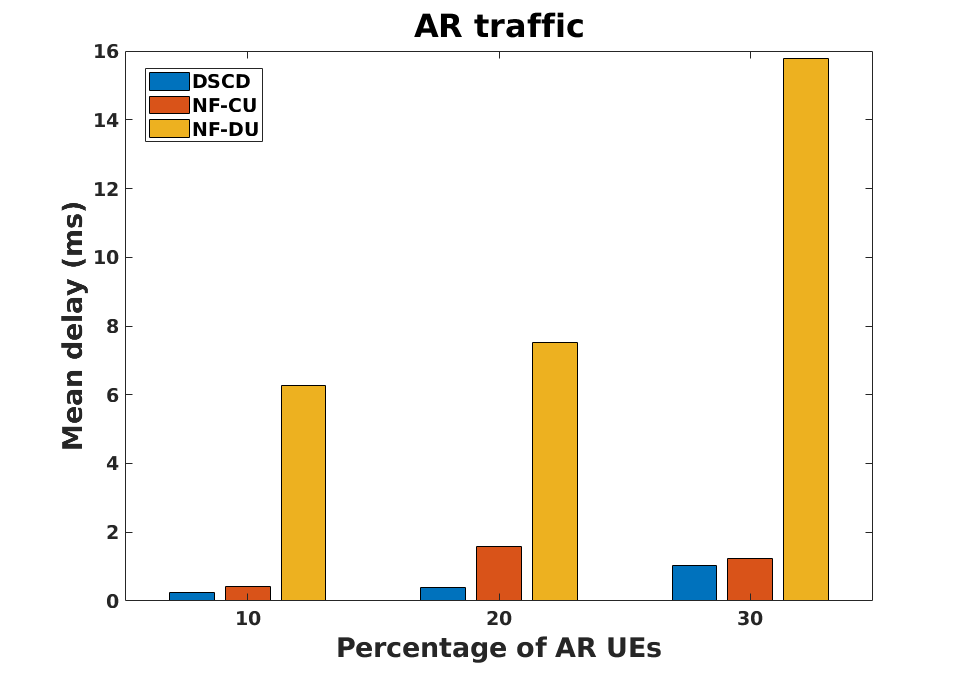}\hfill
  \includegraphics[width=.5\linewidth]{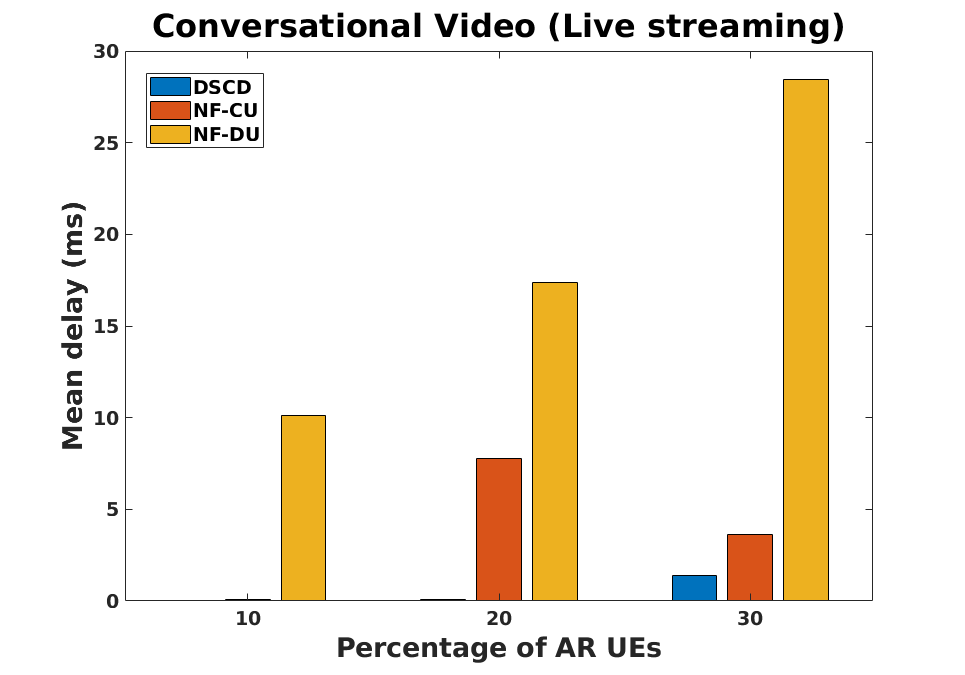}\hfill
  \end{subfigure}\par\medskip
  \caption{
  The scheduling delay for the fixed scenario for DSCD, NF-CU, and NF-DU under AR and Live video stream traffic load.}
  \label{delays1}
\end{figure}

\begin{figure}[t]
  \begin{subfigure}{\linewidth}
  \includegraphics[width=.5\linewidth]{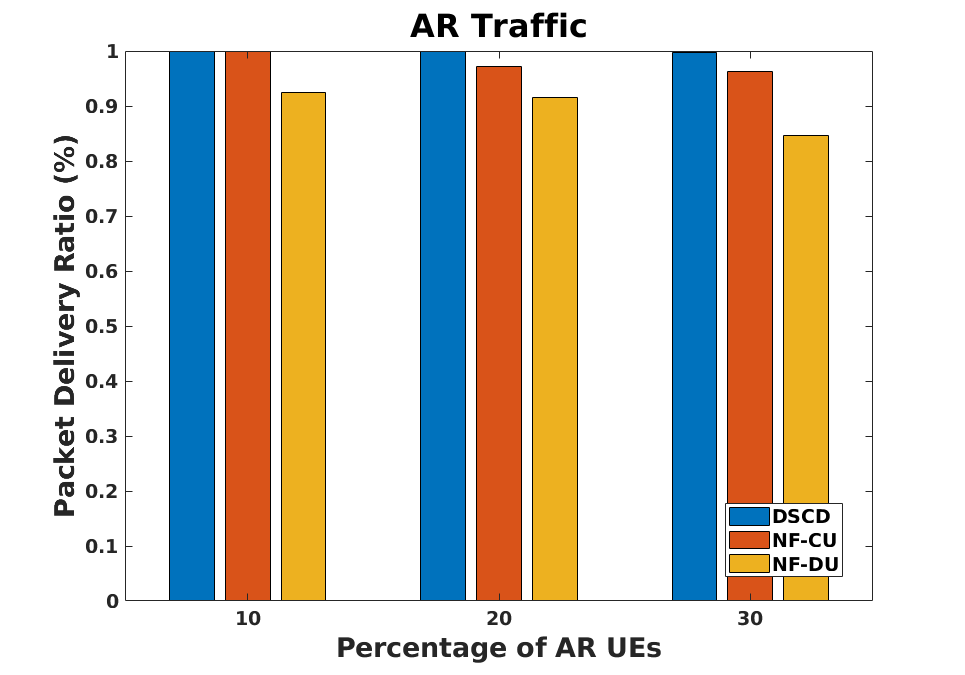}\hfill
  \includegraphics[width=.5\linewidth]{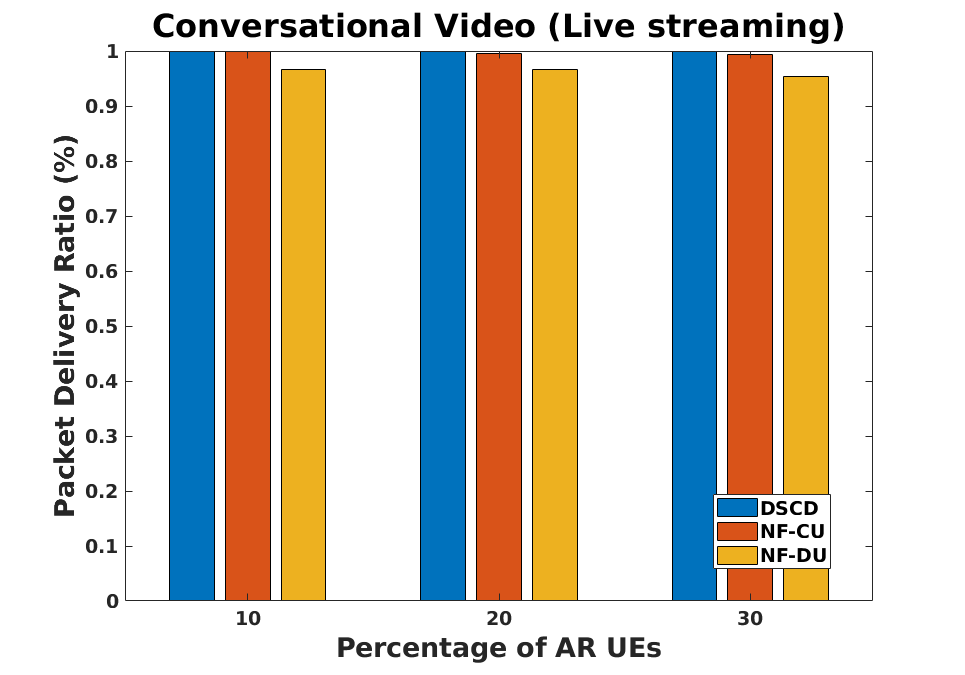}\hfill
  \end{subfigure}\par\medskip
  \caption{
  The packet delivery ratio for the fixed scenario for DSCD, NF-CU, and NF-DU under AR and Live video stream traffic load.}
  \label{delivery1}
\end{figure}

\begin{figure}[t]
  \begin{subfigure}{\linewidth}
  \includegraphics[width=.5\linewidth]{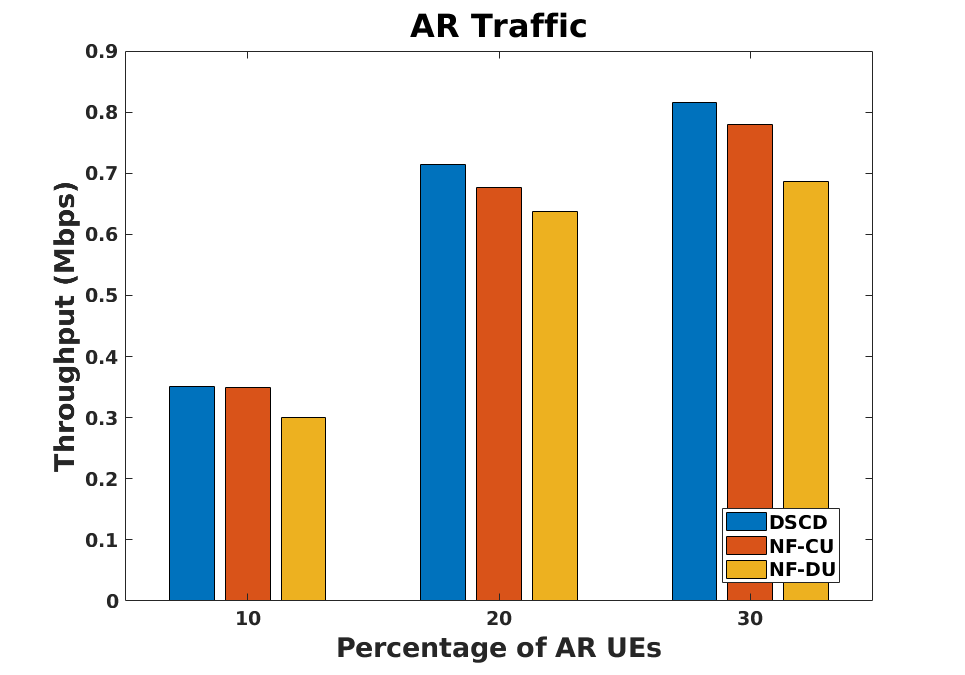}\hfill
  \includegraphics[width=.5\linewidth]{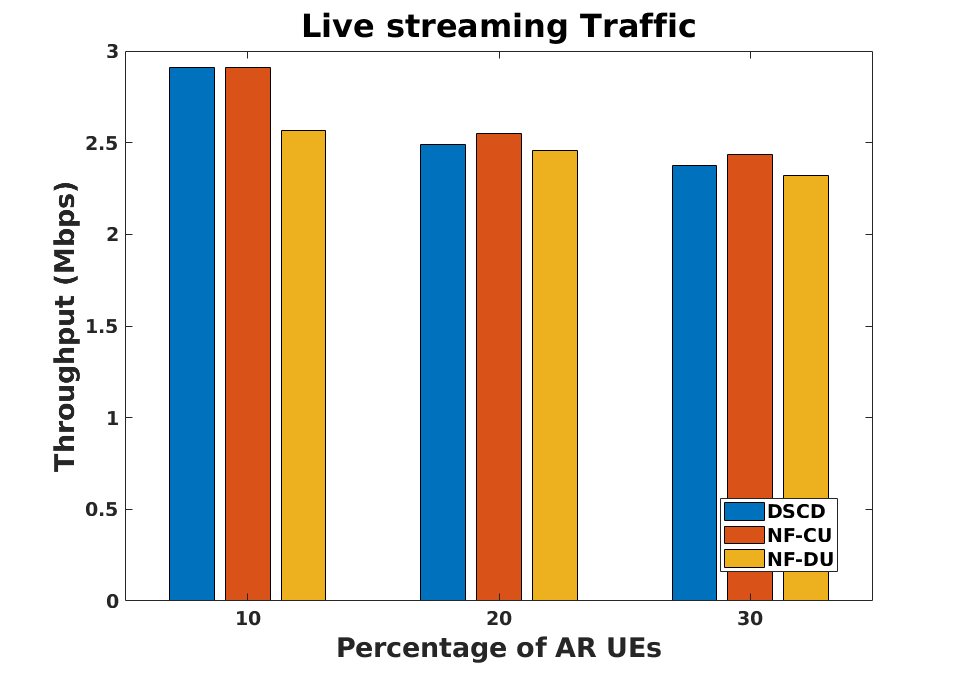}\hfill
  \end{subfigure}\par\medskip
  \caption{
  The mean throughput for the fixed scenario for DSCD, NF-CU, and NF-DU under AR and Live video stream traffic load.}
  \label{throughput1}
\end{figure}

\subsection{Fixed Scenario}


We evaluated two traffic types in the first scenario, including live stream video and virtual reality (VR), with various network loads and when the UEs have no mobility. 

To further elaborate on the results, we show how NFs are dynamically relocated when our proposed technique is used, in Fig. \ref{loadfix}.  In DSCD, unlike the other two models, where the NF is constantly kept at the CU or DU, the NF is dynamically executed at different layers of O-RAN.

Moreover, in Fig. \ref{delays1} and Fig. \ref{delivery1}, the amount of HoL delay and packet delivery ratio with respect to different traffic types with various loads are shown. As we can see, the DSCD can significantly reduce the HoL delay and enhance the packet delivery ratio with respect to NF-DU and NF-CU, respectively. Although executing an NF at DU can be useful for allocating RB to UEs with URLLC traffic, due to the limited observation space of agents compared to the ones located at CU, RBs may not be allocated in a way that the possible inter-cell interference can be prevented. Similarly, If we relocate the NF to CU, although in CU, the NF can access other DUs resource block maps, which can lead to more accurate actions, delay-sensitive packets may not be delivered on time, which can lead to higher HoL and lower packet delivery ratio. Note that we consider HoL delay only. End-to-end delay may have different behavior since one needs to account for CU's distance to the users.

Consequently, as shown in Fig. \ref{throughput1}, the DSCD model has higher throughput than NF-DU and NF-CU. As it can be seen, the overall AR throughput is improved up to 23\%, while the live streaming video traffic throughput is maintained.



\subsection{Mobile Scenario}
To examine the flexibility and adaptability of the proposed model, we consider mobility as well. Here is assumed mobile UEs are vehicles with V2X traffic loads. We also increase the ratio of mobile vehicles from 10\% to 30\% to monitor the DSCD model's performance in different conditions. 

\begin{figure}[t]
  \begin{subfigure}{\linewidth}
  \includegraphics[width=.5\linewidth]{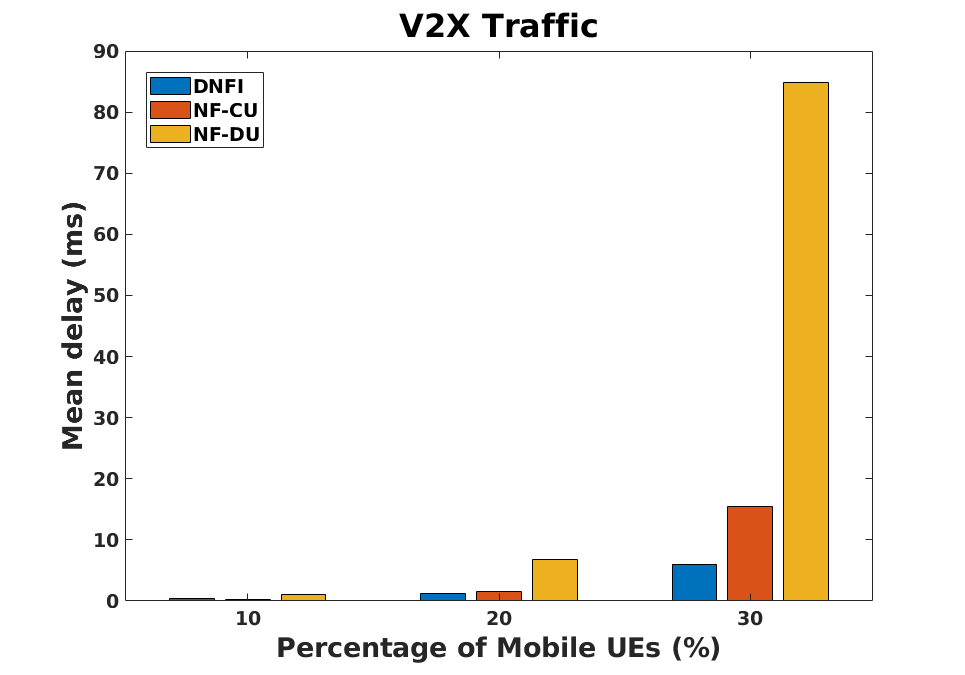}\hfill
  \includegraphics[width=.5\linewidth]{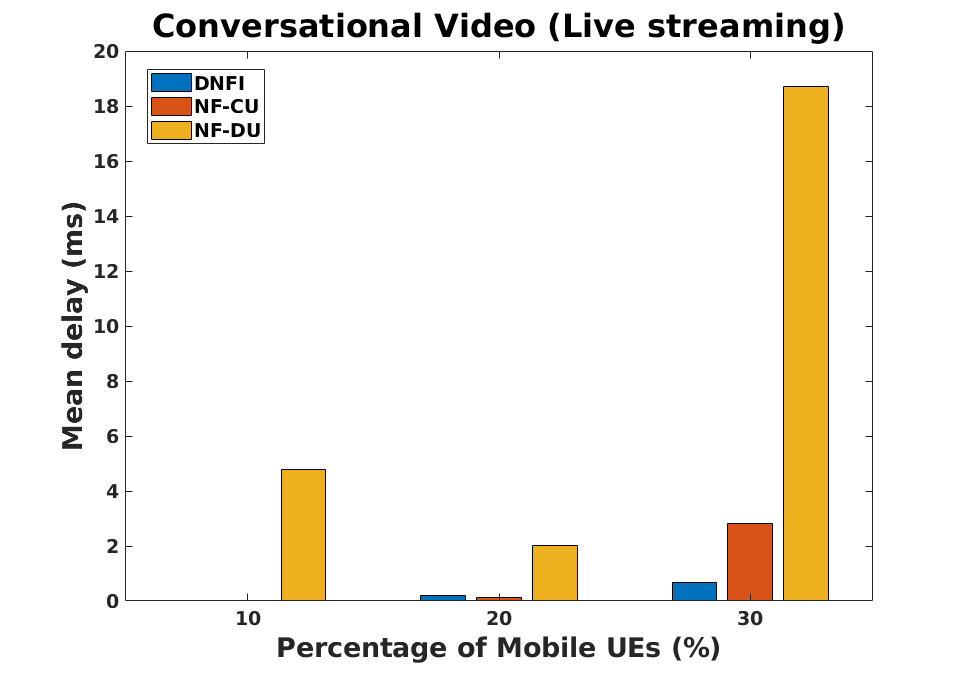}\hfill
  \end{subfigure}\par\medskip
  \caption{The scheduling delay for the mobile scenario for DSCD, NF-CU, and NF-DU under AR and Live video stream traffic load.}
  \label{delays2}
\end{figure}
\begin{figure}[h]
  \begin{subfigure}{\linewidth}
  \includegraphics[width=.5\linewidth]{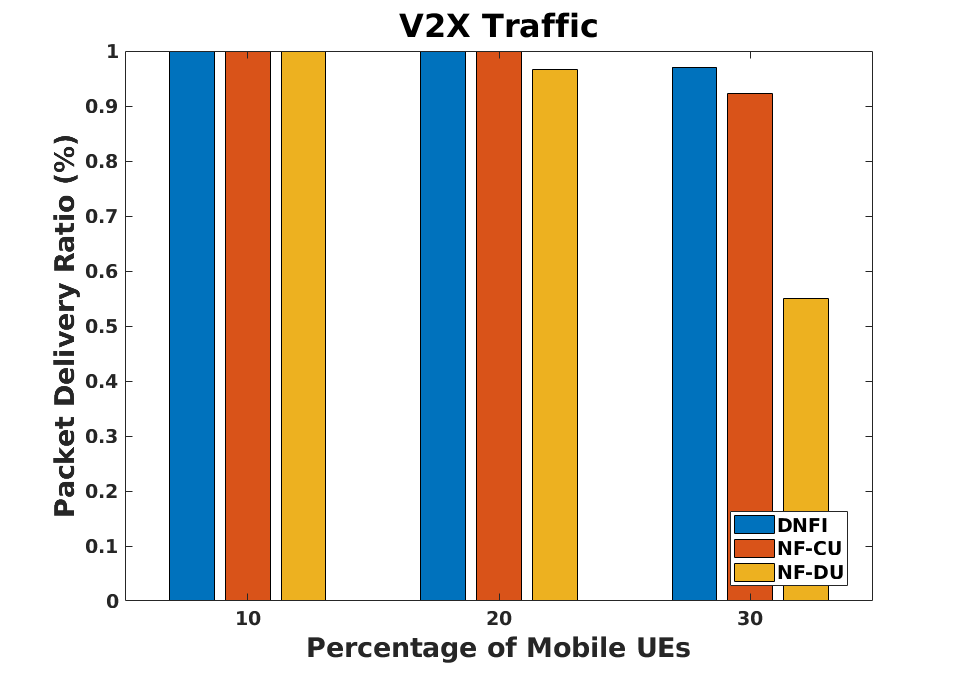}\hfill
  \includegraphics[width=.5\linewidth]{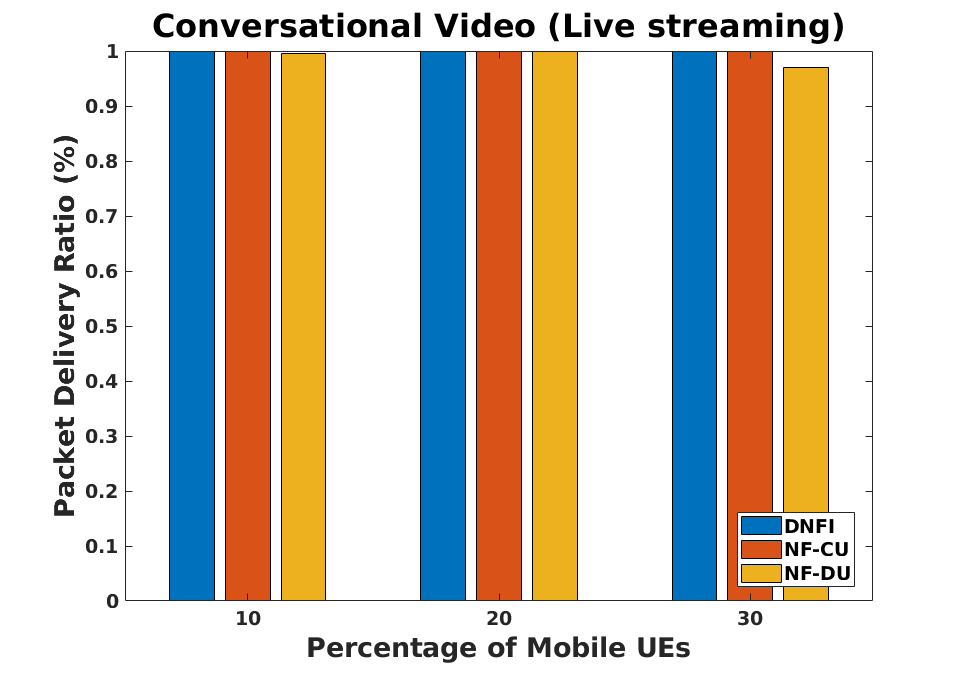}\hfill
  \end{subfigure}\par\medskip
  \caption{The packet delivery ratio for the mobile scenario for DSCD, NF-CU, and NF-DU under AR and Live video stream traffic load.}
  \label{delivery2}
\end{figure}

Additionally, as we can see in Fig. \ref{delays2}, the proposed model can reduce the HoL delay significantly in comparison with the NF-DU and NF-CU by optimizing RB allocation for different traffic types among mobile and fix UEs. As it is depicted, the mean HoL delay for AR and live stream traffics is reduced to about 75 ms and 18 ms, respectively. Likewise, as it is shown in Fig. \ref{delivery2}, the amount of packet delivery ratio in DSCD for video streaming traffic is almost maintained, and in the case of high mobility ratio, it is enhanced noticeably.

\begin{figure}[h]
  \begin{subfigure}{\linewidth}
  \includegraphics[width=.5\linewidth]{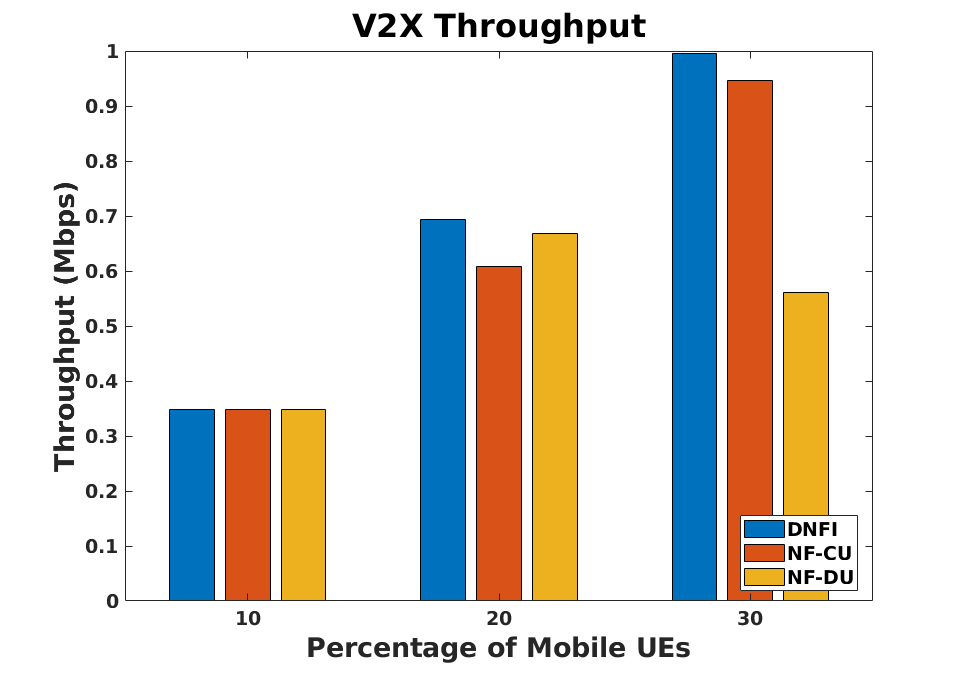}\hfill
  \includegraphics[width=.5\linewidth]{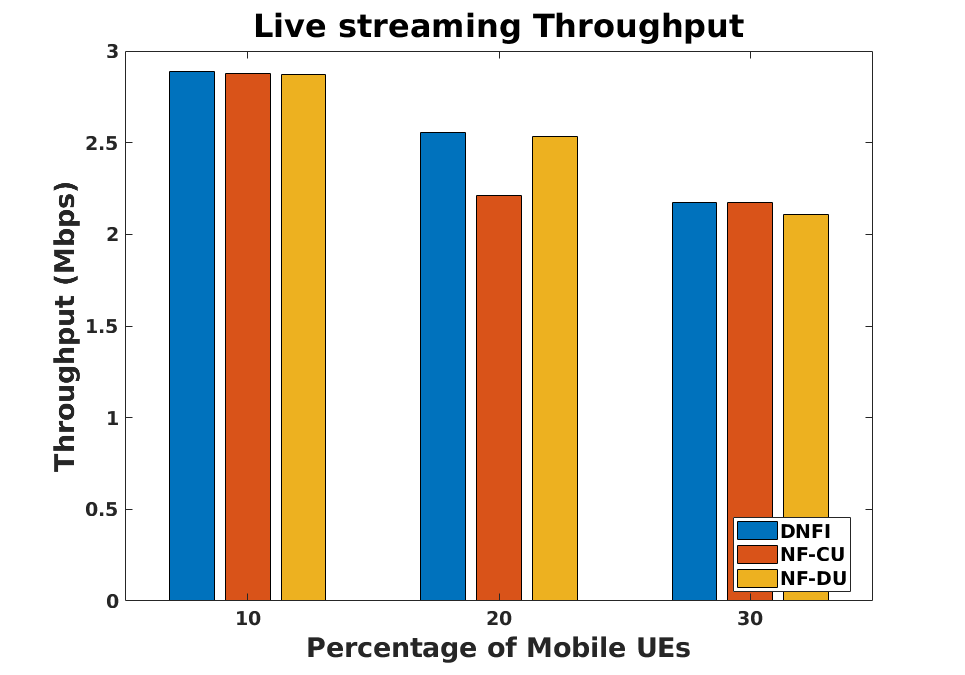}\hfill
  \end{subfigure}\par\medskip
  \caption{The mean throughput for the mobile scenario for DSCD, NF-CU, and NF-DU under AR and Live video stream traffic load.}
  \label{throughput2}
\end{figure}

 We also evaluate the throughput of the DSCD model with respect to the NF-DU and NF-CU, respectively. As shown in Fig. \ref{throughput2}, the DSCD's higher delivery ratio doubled the throughput of the V2X packet, while the overall throughput for video streaming is maintained with respect to NF-DU and NF-CU.    

\section{Conclusion}
In this paper, we focus on the significance of the level of observation in O-RAN for NFs. As an example NF, we focus on resource allocation. We propose a nested reinforcement learning-based technique that involves two RL models working together. The first layer is intelligently deciding on the selection of where to locate the NF in order to hit the balance between best observability and the best performance. Meanwhile, the second layer refines the decisions that are specific to the NF, which in our case, is resource block allocation decisions. Our simulation results show that the proposed dynamic CU-DU selection technique along with the intelligent resource allocation scheme can significantly improve latency, packet delivery ratio, and throughput with respect to NF-DU and NF-CU where the NF is solely deployed at DU or CU, respectively.

\section{Acknowledgement}
This work is supported by Ontario Centers of Excellence (OCE) 5G ENCQOR program and Ciena.

\bibliographystyle{ieeetr}


\bibliography{main}

\end{document}